\documentclass[12pt,twoside]{article}
\usepackage{latexsym}
\usepackage{longtable}
\usepackage{epsfig}
\usepackage{graphicx,psfrag}
\usepackage{amssymb,amsmath,amsthm,booktabs,mathtools}
\usepackage{bm}
\usepackage{color}
\usepackage{subfig}
\usepackage{dutchcal}
\usepackage{yfonts}

\usepackage{hyperref}
\hypersetup{
    colorlinks,
    citecolor=red,
    linkcolor=blue
}

\newcommand{\bc}{\begin{center}}
\newcommand{\ec}{\end{center}}
\def\ba#1{\begin{array}{#1}\displaystyle}
\newcommand{\ea}{\end{array}}

\newcommand{\beq}{\begin{equation}}
\newcommand{\eeq}{\end{equation}}
\newcommand{\beqa}{\begin{eqnarray}}
\newcommand{\eeqa}{\end{eqnarray}}

\newcommand{\n}{\nonumber\\}
\newcommand{\bi}{\begin{itemize}}
\newcommand{\ei}{\end{itemize}}

\def\t#1{\tilde{#1}}
\def\h#1{\hat{#1}}
\def\b#1{\bar{#1}}
\def\frc#1#2{\frac{#1}{#2}}

\newcommand{\p}{\partial}

\newcommand{\bra}{\langle}
\newcommand{\ket}{\rangle}

\newcommand{\R}{{\mathbb{R}}}

\newcommand{\ep}{\epsilon}
\newcommand{\varep}{\varepsilon}

\newcommand{\ri}{{\rm i}}

\newcommand{\dd}{{\rm d}}

\begin{document}

\begin{center}
{\Large {\bf $T{\overline T}$ deformations and\\[0.2cm]  the width of fundamental particles}}

\vspace{1cm}

{\large John Cardy$^{\sharp,\flat}$ and Benjamin Doyon$^*$}
\vspace{0.2cm}

{\small\em
$^\sharp$ Department of Physics, University of California, Berkeley CA 94720, USA
\\
$^\flat$ All Souls College, Oxford OX1 4AL, UK \\
$^*$ Department of Mathematics, King's College London, Strand, London WC2R 2LS, U.K.}
\end{center}
\vspace{1cm}

\noindent We provide a simple geometric meaning for deformations of so-called $T{\overline T}$ type in relativistic and non-relativistic systems. Deformations by the cross products of energy and momentum currents in integrable quantum field theories are known to modify the thermodynamic Bethe ansatz equations by a ``CDD factor". In turn, CDD factors may be interpreted as additional, fixed shifts incurred in scattering processes: a finite width added to the fundamental particles (or, if negative, to the free space between them). We suggest that this physical effect is a universal way of understanding $T{\overline T}$ deformations, both in classical and quantum systems. We first show this in non-relativistic systems, with particle conservation and translation invariance, using the deformation formed out of the densities and currents of particles and momentum. This holds at the level of the equations of motion, and for any interaction potential, integrable or not. We then argue, and show by similar techniques in free relativistic particle systems, that $T\overline T$ deformations of relativistic systems produce the equivalent phenomenon, accounting for length contractions. We also show that, in both the relativistic and non-relativistic cases, the width of particles is equivalent to  a state-dependent change of metric, where the distance function  discounts the particles' widths, or counts the additional free space. This generalises and explains the known field-dependent coordinate change describing $T\overline T$ deformations. The results connect such deformations with generalised hydrodynamics, where the relations between scattering shifts, widths of particles and state-dependent changes of metric have been established.
\vspace{0.5cm}

{\ }\hfill
\today

\newpage

\tableofcontents

\section{Introduction}

In recent years, there has been considerable interest in understanding the physical meaning of the so-called $T{\overline T}$ deformations in 1+1-dimensional quantum field theory, first introduced in \cite{Zam1}. It has been understood \cite{Zam2, Gui, JCNR} that deformations of the Hamiltonian by operators of the form $\int \dd x\, (q(x) \t j(x) - \t q(x) j(x))$, where $q(x),\,\t q(x)$ are conserved densities and $j(x),\,\t j(x)$ are their associated currents, share a number of striking properties. They are perturbatively marginal or non-renormalisable, yet generate models with well controlled UV behaviours \cite{Zam2}. They may be reproduced by appropriate field-dependent changes of coordinates \cite{Conti1, Conti2, JCCF}, with deep connection to quantum gravity \cite{DubJT}. As deformations of QFTs, it has been understood \cite{Tat1,Dub2, Zam2} that they correspond to the adjunction of so-called CDD factors: these are fundamental ambiguities in the analytic $S$-matrix program of integrable QFTs \cite{CDD}. For instance, the $T{\overline T}$ deformation changes the scattering phase $S_{ab}(\theta) = e^{\ri\Phi_{ab}(\theta)}$, where $\theta$ is the rapidity difference in a two-body process, as
\beq\label{changeSintro}
	S_{ab}(\theta) \to S_{ab}'(\theta) = e^{\ri\Phi_{ab}(\theta) - \ri \lambda_{\rm R}M_aM_b\sinh(\theta)}
\eeq
where $M_{a,b}$ are the particles masses and $\lambda_{\rm R}$ is the deformation parameter, with a similar relation for multi-particle processes \cite{Dub2}.

Independently, in the context of integrable one-dimensional systems, a new hydrodynamic theory -- dubbed generalised hydrodynamics (GHD) -- has been developed \cite{GHD1a,GHD1b}, see the review \cite{DoyonLecture2020}. This is a hydrodynamic theory that accounts for the infinite-dimensional manifold of maximal entropy states (the generalised Gibbs ensembles) allowed by the infinite number of conserved quantities \cite{GGE1,GGE2}, see the reviews \cite{GGErev1,GGErev2}. The GHD equations are based on the thermodynamic Bethe ansatz (TBA), and are conservation laws for asymptotic-particle phase-space densities. Interestingly, even in quantum systems, as found in \cite{DYC18}, these conservation laws can be interpreted as kinetic equations for fluids of classical solitons \cite{El-2003,El-Kamchatnov-2005,El2011,CDE16}, where soliton scattering shifts are identified with the TBA differential scattering phase \cite{DYC18},
\beq\label{diffscat}
	\varphi = \frc{d \Phi}{dp}.
\eeq
It has been understood that the GHD Euler-scale equations can also be reproduced by a state-dependent change of coordinates from the Liouville equations for phase-space densities \cite{DSYgeom}, see also \cite[Sect 4.3]{DoyonLecture2020}. The Liouville equations are the hydrodynamic equations for free particles, and the state-dependent change of coordinates changes the definition of space by attributing a nonzero, generically momentum-dependent effective widths to the otherwise free particles. In this interpretation, these widths are what account for the soliton scattering shifts \cite{DYC18,DSYgeom}.

The special case where the width is not momentum dependent is of particular interest: for positive widths, this is the case of the gas of hard rods \cite{Aizenman1974ergodic,Boldrighini1983,Spohn-book}. The hard rod system is a system of particles with equal, finite widths $\lambda>0$, subject to elastic collisions and otherwise free. In this case, the technique of collapsing the system to free particles is well known, for instance at the basis of the rigorous derivation of its hydrodynamics in \cite{Boldrighini1983}; its geometric interpretation was given in \cite{Doyon-Spohn-HR-DomainWall}. The thermodynamics and hydrodynamics of the hard rod fluid is based on TBA equations where $\varphi=-\lambda$ is independent of momentum \cite{DYC18} (that is, coming from a pure CDD factor).

Combining \eqref{changeSintro} with \eqref{diffscat}, under a $T{\overline T}$ deformation, the differential scattering phase changes as $\varphi\to\varphi' = \varphi + \lambda$. Interpreting $\varphi$ as a scattering shift, and in turn, following the GHD intuition, as a particle width, one may seek to give $T{\overline T}$ deformations the physical interpretation as elongations of the widths of the fundamental particles of the theory.

The goal of this paper is to put this remark on firm ground, and to suggest its universality. In order to do so, we first analyse the {\em non-relativistic version} of the $T{\overline T}$ deformation, the so-called $qj$ deformation. We consider a system formed out of $N$ particles in one dimension, quantum or classical, with arbitrary dispersion relation and interaction. An explicit evaluation of the effect of the $qj$ deformation shows that the flow generated by the deformation has the simple effect of changing the system into a system of $N$ finite-width particles, with the same interaction and dispersion relation. We treat separately both cases where the un-deformed (and thus also the deformed) particles have a hard core, exchanging their momenta at collisions, and where they go through each other at collisions. In the case of hard-core but otherwise non-interacting particles, this reproduces the well known system of hard rods. The effect is valid for any interaction, and does not require integrability. We prove this effect at the level of the equations of motion, hence it holds microscopically, not just in the thermodynamics or hydrodynamic limits. Then, we show how the relativistic $T\overline T$ deformation has a similar effect, accounting for length contractions.

The paper is organised as follows. In section \ref{sectnr} we describe the general system of non-relativistic particles considered, and express and prove our main results in this context. In section \ref{sectint}, we provide interpretations, connecting  with concepts developed in the $T\overline T$ and GHD literature. In particular, we describe the connection to the free hard rod model and to state-dependent changes of metric, and we study the deformation of the scattering phase, of the hydrodynamics and of the thermodynamics of the model. In section \ref{sectrel}, we explain how the results and interpretations naturally generalise to relativistic systems. We finally conclude in Section \ref{sectcon}.

\section{Non-relativistic case}\label{sectnr}

\subsection{The setup: interacting particles}\label{ssectinteracting}

Consider a system, be it quantum or classical, composed of $N$ particles, taken for convenience to be identical, lying in one dimension of space in infinite volume (that is, lying on the line). The Hamiltonian is assumed to be particle-preserving, and to separate into a kinetic and potential term,
\beq\label{H}
	H = \sum_{n=1}^{N} \omega(p_n) +V(x_1,\ldots,x_N),
\eeq
where $p_n$ and $x_n$ are canonically conjugate momentum and position variables.

The dispersion relation $\omega(p)$ is kept arbitrary. A natural setup is where there is Galilean invariance, in which case it takes the usual form
\beq
	\omega(p) = \frc{p^2}{2M}\qquad\mbox{(Galilean case),}
\eeq
$M$ being the mass of the particles; however it is not necessary to assume this here. In general, we assume the potential to be translation invariant,
\beq\label{translationinv}
	V(x_1,\ldots,x_N) = V(x_1+a,\ldots,x_N+a).
\eeq

We will discuss two separate physical situations, or ``pictures": the {\em hard-core picture}, where the particles have a hard core, and exchange their momenta at two-body collisions; and the {\em go-through picture}, where particles go through at two-body collisions. In both cases this is in addition to the interaction represented by the potential $V(x_1,\ldots,x_N)$.

In the hard-core picture, one may just take the potential to satisfy
\beq\label{hc}
	V(x_1,\ldots,x_N) = \infty \quad \mbox{if}\quad
	x_n\geq x_m \mbox{ for some $n<m$
	(hard-core picture).}
\eeq
For instance, this includes the standard form of a two-body interaction,
\[
	V(x_1,\ldots,x_N) = \sum_{n>m} V(x_n-x_m),
\]
with a hard-cord potential $V(x)=\infty$ for $x<0$. As, with a hard core, the particles cannot cross each other, the ordering
\beq\label{or}
	x_{n+1}>x_{n},\quad n=1,2,\ldots N-1
	\quad \mbox{(hard-core picture)}
\eeq
is kept throughout the dynamics. In the classical case, this is a reduction of the phase space accessible, and in the quantum case, of the support of the many-body wave function.

By translation invariance, one may see the potential as a function of the distances between pairs of particles, this function becoming infinite whenever one of the distances is non-positive:
\beq\label{Vdnm}
	V(x_1,\ldots,x_N) = V(\{d_{nm}:n>m\}),\qquad d_{nm} = x_n-x_m \quad \mbox{(hard-core picture).}
\eeq

The hard-core picture will be useful in order to derive the microscopic effect of the deformation we will study.

In the go-through picture, instead of \eqref{hc}, we impose the condition that the potential be permutation-symmetric,
\beq\label{ps}
	V(x_1,\ldots,x_N) = V(x_{\sigma_1},\ldots,x_{\sigma_N})
	\quad \mbox{if} \quad i\mapsto \sigma_i\mbox{ is a permutation (go-through picture)}.
\eeq
In the classical setting, it is clear that the system \eqref{H} with a potential satisfying \eqref{ps} can be mapped to that with a potential that has the same form in the region of phase space where the ordering \eqref{or} holds, and that otherwise satisfies the hard-core condition \eqref{hc}. Indeed, thanks to the elastic collision of hard-core particles, where momenta are exchanged, the trajectories under both dynamics can be mapped to each other by a permutation. Permutation elements are applied sequentially following the order of particle crossings, resp.~collisions (in the dynamics with \eqref{ps}, resp.~\eqref{hc}). A similar argument can be made for indistinguishable particles in the quantum case, with either Bosonic or Fermionic statistics. The relation of this go-through picture to the hard-core picture is that the go-through trajectories are the trajectories of the ``local velocity tracers" in the hard-core picture. That is, the trajectory of a local velocity tracer is formed of parts of trajectories of hard-core particles, in such a way that on the local velocity tracer's trajectory, the velocity varies smoothly; this is consistent as collisions of hard core particles are elastic, velocities being instantaneously exchanged.

We will use this mapping between trajectories in order to deduce the microscopic dynamics after deformation in the go-through picture, from that in the hard-core picture. Because of this mapping, the two pictures are simply related, and most of the discussions below, for indistinguishable particles, apply equally well to both.

We discuss in Section \ref{ssectmanyspecies} the case of particles having different species, instead of being indistinguishable. In this case, the two pictures are not so simply related; we will be able to derive the deformed microscopic dynamics in the hard-core picture, but will only conjecture that for the go-through picture.

Below we will also discuss results on the deformed scattering of particles, which can be deduced from the deformed microscopic dynamics. In these discussions, we will assume that the potential is such that the particles are the correct asymptotic objects: that there are no stable asymptotic bound states. This is in order to simplify the discussion. For the scattering of asymptotic particles, our results are in fact stronger: by the techniques we introduce, the deformation can be evaluated explicitly in both hard-core and go-through pictures, and both for indistinguishable particles and for many species.

The system may admit a number of conserved quantities; it may or may not be integrable. But by particle number conservation and invariance under space translation, it admits at least two: the number of particles $Q_0 = N$ and the momentum $Q_1 = P = \sum_n p_n$. From these, one may define densities $q_i(x,t)$, with $Q_i = \int \dd x\,q_i(x,t)$:
\beq\label{q0q1}
	q_0(x,t) = \sum_n \delta(x_n(t)-x),\quad q_1(x,t) = \sum_n \delta(x_n(t)-x)p_n(t).
\eeq
In the quantum case, there is an operator-ordering ambiguity in the definition of $q_1(x,t)$; however this only leads to total space derivatives, a gauge transformation of the charge density which does not change the total charge $Q_1$, and neither which affect our main result. It is a simple matter to verify that local conservation laws are satisfied:
\beq\label{cons}
	\p_t q_i(x,t) + \p_x j_i(x,t) = 0
\eeq
for appropriate $j_i(x,t)$. In particular, if the model is Galilean invariant, then $j_0(x,t) = \t q_1(x,t)$ with $\t q_1(x,t)$ a local momentum density\footnote{Explicitly, $\t q_1(x,t)  = \frc12 \sum_n (\delta(x_n(t)-x)p_n(t) + p_n(t) \delta(x_n(t)-x))$, which is, in the quantum case, a different gauge choice from that made in \eqref{q0q1}.}; but Galilean invariance is not necessary for our results to hold.

\subsection{Main result}\label{sectmain}

We wish to consider a deformation of the Hamiltonian which naturally adapts the widely studied $T{\overline T}$-deformation \cite{Zam1} to non-relativistic systems. As in the non-relativistic limit the system preserves the number of particles and is translation invariant, the above is a natural setup for such considerations. This is essentially equivalent to taking the non-relativistic limit of the relativistic energy in which the rest mass dominates. Since the additional non-relativistic energy is then separately conserved, it is also possible to consider a different deformation whereby the two conserved currents are the non-relativistic energy and the momentum. This was treated in an earlier paper \cite{JCNR}. However this is not in general as solvable because the space component of the energy current does not have a simple general form.  

 Then, a natural choice of the deformation is the infinitesimal change $H^{(\lambda)}\to H^{(\lambda+\delta\lambda)} = H^{(\lambda)}+\delta\lambda\Delta^{(\lambda)} + O((\delta\lambda)^2)$ with 
\beq\label{Delta}
	\Delta^{(\lambda)} = \int \dd x\,(q^{(\lambda)}_0(x-\varep)j^{(\lambda)}_1(x)- j^{(\lambda)}_{0}(x-\varep)q^{(\lambda)}_1(x) )
\eeq
where $\varep>0$ is an infinitesimal point-splitting regularisation. By the same arguments as those expressed in \cite{Zam2}, all Hamiltonians $H^{(\lambda)}$ along the flow generated by this infinitesimal change must preserve (a form of) translation invariance and the number of particles; thus there are charges $Q^{(\lambda)}_i$ for $i=0,1$ throughout the flow, and at every $\lambda$, the deformation is \eqref{Delta} with respect to the associated charge densities.

In the hard-core picture, where particles are originally hard-core points, our main results are as follows:
\bi
\item For $\lambda>0$, the Hamiltonian $H^{(\lambda)}$ is a system where {\em all particles acquire a hard core with a positive width} $\lambda$. The particles are therefore hard rods, with the coordinates $x_n$ being the leftmost boundaries of the rods. The same interaction potential \eqref{Vdnm} applies, with $d_{nm}$  the physical distance between the rods $n$ and $m$ (not between the coordinates $x_n$ and $x_m$).
\item For $\lambda<0$, the Hamiltonian $H^{(\lambda)}$ is a system where {\em free space has been enlarged}, with an additional distance $-\lambda$ between each particle pair. The particles may be seen as  negative-width rods: with $x_n$ being the leftmost boundaries of the rods, this means that these leftmost boundaries can go pass each other and then travel up to a distance of $-\lambda$ before the hard cores come into contact. The same interaction potential \eqref{Vdnm} applies, again with $d_{nm}$ the physical distance between the (negative-width) rods $n$ and $m$.
\ei

In the go-through picture, where original particles go through each other at collisions and \eqref{ps} is assumed, we may simply apply the mapping described after Eq.~\eqref{ps} to the above result in order to describe the effects of the deformation. Recall that the go-through trajectories are the trajectories of the local velocity tracers in the hard-core picture. Under this mapping, the results above also hold in the go-through picture. Their interpretation is therefore that, in addition to the change in the potential itself due to the deformation, for $\lambda>0$, local velocity tracers in a pair {\em jump forward} a length $\lambda$ when they reach a distance $\lambda$ from each other, while for $\lambda<0$, they {\em jump backward} a length $-\lambda$ when they have passed each other a distance $-\lambda$.

In order to show these results, it is sufficient to consider the hard-core picture, as by the argument above, the results in the go-through picture can be deduced by the mapping on trajectories. We note that the flow generated by the deformation \eqref{Delta} is in fact the result of a canonical transformation with generator \cite{GEN,kruthoff2020}
\beq\label{Xgen}
	X = -\int \dd y\,\int_{-\infty}^{y-\varep} \dd x\,q_i(x)q_j(y)
\eeq
with $i=0$, $j=1$. Indeed, in general, with conserved densities satisfying \eqref{cons}, we have
\beqa
	\{H,X\} &=& -\int \dd y\int_{-\infty}^{y-\varep}\dd x\,\{H,q_i(x)q_j(y)\}\n
	&=& \int \dd y\int_{-\infty}^{y-\varep}\dd x\,(\dot q_i(x)q_j(y) + q_i(x)\dot q_j(y))\n
	&=& -\int \dd y\int_{-\infty}^{y-\varep}\dd x\,(\p_x j_i(x)q_j(y) + q_i(x)\p_y j_j(y))\n
	&=& \int \dd y\, (q_i(y-\varep) j_j(y) - j_i(y-\varep)q_j(y))
\eeqa
where $\{\cdot,\cdot\}$ is either the classical Poisson bracket, in the classical case, or the commutator $-\ri [\cdot,\cdot]$, in the quantum case (we set $\hbar=1$). The full deformation of the Hamiltonian is the result of the flow $\lambda \mapsto H^{(\lambda)}$ generated by $X$,
\beq\label{flownr}
	\frc{d H^{(\lambda)}}{d\lambda} = \{H^{(\lambda)},X\}.
\eeq
The generator $X$ may be explicitly worked out:
\beqa
	X &=& -\sum_{n,m} \int \dd y\int_{-\infty}^{y-\varep}\dd x\,\delta(x-x_n) \delta(y-x_m) p_m\n
	&=& -\sum_{n,m} \int_{-\infty}^{x_m-\varep}\dd x\,\delta(x-x_n)p_m  = -\sum_{n,m} \Theta(x_m-\varep-x_n)p_m  = -\sum_{m>n} p_m  \n
	&=& -\sum_m (m-1) p_m .\label{X}
\eeqa
In the fourth step we used the fact that the particles are ordered, Eq.~\eqref{or}, and that $\varep$ is a positive infinitesimal. The result is valid both for $X$ as a classical phase-space function, or as an operator on the $N$-particle Hilbert space.

The results then immediately follow from applying the generator to the canonical coordinates, $\{x_n,X\} = -(n-1)$, $\{p_n,X\} = 0$. Clearly, the formal solution is
\beq\label{transfo}
	p_n^{(\lambda)} = p_n,\qquad x_n^{(\lambda)} = x_n - (n-1)\lambda.
\eeq
The flow is obtained by taking the Hamiltonian $H$ on the coordinates $x_n^{(\lambda)}$ and $p_n^{(\lambda)}$ thus generated:
\beq\label{HHl}
	H^{(\lambda)}(p_\bullet,x_\bullet) = H(p^{(\lambda)}_\bullet,x^{(\lambda)}_\bullet).
\eeq
Therefore,
\beq\label{Hlambda}
	H^{(\lambda)} = \sum_{n=1}^{N} \omega(p_n) +V^{(\lambda)}(x_1,\ldots,x_N),\quad
	V^{(\lambda)}(x_1,\ldots,x_N) = V(x_1,x_2-\lambda,\ldots,x_N-(N-1)\lambda).
\eeq
Equivalently, the phase-space trajectories generated by $H^{(\lambda)}$ are  $\big(x_n^{(-\lambda)}(t),p_n^{(-\lambda)}(t)\big)$, if those generated by $H$ are $(x_n(t),p_n(t))$. 

The transformation can also be explicitly written in quantum systems in terms of the wave function: we may use the flow equation $\psi^{(\lambda)} =  e^{\ri \lambda \h X}\psi$ along with \eqref{X}, $\h X = -\sum_m (m-1) \h p_m$, obtaining:
\beq\label{psilambda}
	\psi^{(\lambda)}(x_1,\ldots,x_N) =
	\psi(x_1,x_2-\lambda,\ldots,x_N-(N-1)\lambda).
\eeq

In particular, if $\psi$ is a Hamiltonian (generalised) eigenfunction of energy $E$, then $H^{(\lambda)} \psi^{(\lambda)} = E\psi^{(\lambda)}$: the energies are unchanged along the flow. This is valid for a system on infinite volume. If the system is on finite volumes, then, say for $\lambda>0$, as the particles gain length, energies are equal up to a change of the effective volume. Let $\psi$ be an eigenfunction for $H$ of energy $E$, on a volume $R$. The volume is here defined as the range of values that any coordinate can take. Then by \eqref{psilambda}, $\psi^{(\lambda)}$ is an eigenfunction for $H^{(\lambda)}$, but where the range of values of any coordinates is instead $R-\lambda N$. That is,
\beq\label{EE0nonrel}
	E^{(\lambda)}(R-\lambda N) = E^{(0)}(R).
\eeq

This formal solution to the flow needs some clarifications. In the case $\lambda>0$, the ordering of the coordinates $x_n$ is kept throughout the dynamics generated by $H^{(\lambda)}$, as the potential $V^{(\lambda)}(x_1,\ldots,x_N)$ still satisfies the hard-core condition \eqref{hc}; equivalently the trajectories $\big(x_n^{(-\lambda)}(t),p_n^{(-\lambda)}(t)\big)$ keep their ordering. Therefore, the form \eqref{X} of the generator, which uses the ordering \eqref{or}, can be applied in calculating the right-hand side in the flow equation \eqref{flownr}, and the solution \eqref{Hlambda} follows immediately. In the quantum context, this hard-core condition is a non-transmission condition; it remains valid thanks to \eqref{hc}, which forbids tunnelling, thus restricting the space of wave functions to those with ordered arguments.

However, for $\lambda<0$, the strict hard-core condition is {\em broken} under the flow. The meaning of the solution \eqref{Hlambda}, in this case, is via an analytic continuation in $\lambda$: as the analytically continued flow is still generated by \eqref{X} in that form, the solution \eqref{Hlambda} holds throughout.

The solution \eqref{Hlambda} immediately gives the main results expressed above. For instance, the potential $V^{(\lambda)}(x_1,\ldots,x_N)$ is infinite if $x_i>x_{i+1}-\lambda$, thus if the distance $d_{i+1,i}$ is smaller than $\lambda$: for $\lambda>0$,  this is the hard-rod condition, and thus particles have acquired an additional width $\lambda$.

\section{Interpretations}\label{sectint}

\subsection{Free hard rods}\label{ssecthr}

The case of free particles, where $V(x_1,\ldots,x_N)=0$ whenever the ordering \eqref{or} holds, contains already most of the interesting physics within the deformation \eqref{Delta}. The above shows that the flow generated by \eqref{Delta} transforms this into the system of free hard rods of widths $\lambda>0$. The system of classical hard rods has been studied since long ago and is still relevant in current studies of non-equilibrium dynamics \cite{Aizenman1974ergodic,Boldrighini1983,Spohn-book,Doyon-Spohn-HR-DomainWall}. Relating it to a system of free point-like particles, by displacing the rod with label $n$ by a distance $n\lambda$ to its left, is a well-known technique at the basis of many studies. The mapping to free particles is the transformation \eqref{transfo}: the dynamics on the coordinates $x^{(\lambda)}_n$ is indeed that of free particles, as per \eqref{HHl}. Here, we have shown that this transformation is exactly reproduced by the flow generated by \eqref{Delta}; and further, that this applies as well to interacting (quantum and classical) hard rods, with arbitrary translation-invariant interaction potential.

\subsection{State-dependent change of metric}

The notion of enlarging the particles ($\lambda>0$), or enlarging the space in-between them ($\lambda<0$), naturally points to a geometric interpretation. In this interpretation, the flow $\lambda\mapsto H^{(\lambda)}$ corresponds to a special change of metric, that is determined by the presence of the particles.

For illustration, let us consider the case $\lambda>0$, the original system ($\lambda=0$) being composed of point-like particles. Then $H^{(\lambda)}$ is the Hamiltonian for a system of interacting hard rods of widths $\lambda$. Let us fix a point far to the left (say) of all particles, $x_0< x_n$ and $x_0< x_n^{(\lambda)}\;\forall n$. We define a family of metrics, parametrised by $\lambda$, via a family of distances $d^{(\lambda)}(x-x_0)$ to this point with the property that $d^{(\lambda)}(x_n-x_0) = x_n^{(\lambda)}-x_0$. At $\lambda=0$ this is the distance for the usual metric on $\R$, and using \eqref{transfo}, we may set
\beq\label{metric}
	d^{(\lambda)}(x-x_0) = d^{(0)}(x-x_0) - \lambda L(x)
\eeq
where $L(x)$ is the number of particles to the left of the point $x$: $L(x) = \#\{n:x_n<x\}$. Fort $\lambda\neq 0$ this is a distance associated to a {\em state-dependent metric}, which explicitly depends on the particles' configuration.

The statement is that the dynamics $H$ for the point-like particles, but considered on the space with configuration-dependent metric induced by \eqref{metric}, reproduces the dynamics $H^{(\lambda)}$ for the left-most points of the hard rods. The space taken away in \eqref{metric} represents the finite widths of the hard rods. Technically, this follows as putting the distance \eqref{metric} on $H$ amounts to considering the Hamiltonian $H(p_\bullet^{(\lambda)},x_\bullet^{(\lambda)})$; the equality \eqref{HHl} proves the statement. This statement is simply a re-interpretation of \eqref{HHl} with \eqref{transfo}.

The interpretation, including both signs of $\lambda$, is perhaps clearest in the go-through picture, considering the local velocity tracers as discussed around Eq.~\eqref{ps}. The distance $d^{(\lambda)}(x)$ is then to be interpreted as measuring the ``free space" in which the particles (velocity tracers) are allowed to travel, within the real space whose distances are measured by $d^{(0)}(x)$. In the case $\lambda>0$, this free space is smaller, as particles are affected by jumps forward by $\lambda$ when they get too close to each other: in this case, the free space is simply the space between the rods, available for actual travel. The opposite holds for $\lambda<0$, as the jumps are backward: in this case, {\em additional space} is present in which particles can travel.

Changes of metric that encode particles' widths as in \eqref{metric} have been discussed at length in the framework of GHD, see \cite{Doyon-Spohn-HR-DomainWall,DSYgeom}, see also \cite[Sect 4.3]{DoyonLecture2020}. In particular, the above description was given in the hard rod case \cite{Doyon-Spohn-HR-DomainWall}: it is observed that the hydrodynamics of the hard rod gas is obtained from the trivial hydrodynamics of free particles if the metric is modified as per \eqref{metric}. More generally, the notion of a state-dependent change of metric that trivialises the hydrodynamics of integrable systems has been extended beyond the hard rod gas in \cite{DSYgeom}. In this context, the change of metric is determined by the generically momentum-dependent differential scattering phase -- see below for a discussion of scattering. Here, we see that the notion of change of metric -- without momentum dependence -- applies to interacting hard rods as well, without the need for integrability.

The related idea of enlargement of free space appeared first in the context of GHD. In particular the flea gas of \cite{DYC18} provides an explicit microscopic model (or algorithm) where appropriate jumps implement the additional free space in an integrability-preserving way. The quantity of free space is related to the ``density of states", a natural quantity from the TBA formulation, see \cite{DoyonLecture2020} and references therein. The flea gas generalises the concept to momentum-dependent free space. The more recently proposed zigzag model \cite{Donahue2019} is another implementation of additional (momentum-independent) free space. Again, in the present work we see that there is no need for integrability for the notion to emerge consistently.

In the relativistic case, the $T{\overline T}$ deformation has also been associated to a state-dependent coordinate transform in \cite{Conti1,Conti2,JCCF}. It is a simple matter to recover the above result using the arguments made in that context. In relativistic models, one write the deformation of the action using the energy-momentum tensor $T^{ab}$, as (taking again a small point-splitting distance $\varep>0$)
\beq
	\delta S = 2\delta\lambda_{\rm R} \int \dd^2 x\,\ep^{ab}\ep^{cd}T_{ac}(x-\varep)T_{bd}(x).
\eeq
One then regards it as an infinitesimal change of metric $g^{bd} \to g^{bd} + h^{bd}$ with
\beq
	h^{bd}(x) = -2\delta\lambda_{\rm R} \ep^{ab}\ep^{cd}T_{ac}(x-\varep)
	= -2\delta\lambda_{\rm R} \p_{x^d}\Big(\ep^{ab}\int_{-\infty}^{x-\varep}  \dd x'_e\,\ep^{ce}T_{ac}(x')\Big)
\eeq
(where conservation of the stress-energy tensor ensures that the integral is path independent). With $h^{bd} = \p_{x^d}\delta x^b + \p_{x^b}\delta x^d$ for a diffeomorphosm $x^b \to x^b + \delta x^b$, we get
\beq
	\delta x^b = \delta\lambda_{\rm R} \ep^{ab}\int_{-\infty}^{x-\varep}  \dd x'_e\,\ep^{ce}T_{ac}(x').
\eeq
This becomes, with the integral taken on a time slice,
\beq
	\delta x^1 = \delta\lambda_{\rm R}\times\mbox{energy in $(-\infty,x^1-\varep]$},\quad
	\delta x^0 = \delta\lambda_{\rm R}\times\mbox{momentum in $(-\infty,x^1-\varep]$}.
\eeq
Thus, in particular, for $\delta\lambda_{\rm R}>0$ the positions move towards the right by $\delta\lambda_{\rm R}$ times the total quantity of energy to the left. In the non-relativistic limit, the energy in a region is dominated by the total mass of all particles in this region. Setting $\delta \lambda = M\delta\lambda_{\rm R}$ where $M$ is the mass of each particle, we recover the result above.

\subsection{Scattering}

Let us assume that the scattering problem is well defined on the fundamental particles of the undeformed model -- that is, the potential vanishes fast enough at large separations. We wish to analyse the $\lambda$-deformation of the scattering shifts. It is clear, from the physical picture given above, that this should simply be a constant additions to all scattering shifts, corresponding to the additional widths of the particles ($\lambda>0$) or the additional free space ($\lambda<0)$. It is instructive to work this out more precisely.

Consider for simplicity a two-body scattering process; by conservation of momentum and energy, momenta as simply exchanged asymptotically, and it is convenient to trace the individual momenta. Consider the labels 1 and 2 associated to momenta $p_1$ and $p_2$. We assume positions $x_1<x_2$ in the ``in" configuration and $x_1>x_2$ in the ``out" configuration. In the undeformed system, the associated scattering shifts are $\varphi_n$, defined as the shifts of the local velocity tracers after scattering. We take the convention according to which $\varphi_n>0$ ($\varphi_n<0$) for a shift backward (forward) with respect to the side on which the collision happens. For instance,
\beq\label{varphi1}
	\varphi_{1} = x_1^{(\rm in)} -
	x_1^{(\rm out)},\quad
	x_n^{(\rm out,in)} = \lim_{t\to\pm\infty} x_n(t)-v_n^\infty t.
\eeq
Here and below $v_n^\infty = v(p_n(t=\pm\infty))$, with $v_1>v_2$, are the asymptotic velocities, where $v(p) = \omega'(p)$ is the velocity of the particle of momentum $p$. Note that by energy and momentum conservation, for generic dispersion relation $\omega(p)$ the set of momenta is preserved, which we make use here. In quantum systems, this displacement is associated with a scattering angle $\Phi(p_1,p_2)=-\ri \,\log S(p_1,p_2)$ via\footnote{There is no ambiguity in the definition of $\Phi(p_1,p_2)$ as the numerical ordering of momenta, hence of velocities, determine the assumed particle ordering in the ``in" and ``out" configurations.}
\beq\label{scatphase}
	\varphi_{1} = \frc{\p\Phi}{\p p_1}.
\eeq
Likewise,
\beq
	\varphi_{2} = x_2^{(\rm in)}-
	x_2^{(\rm out)} =
	\frc{\p\Phi}{\p p_2}.
\eeq

\begin{figure}
\bc
\includegraphics[width=7cm]{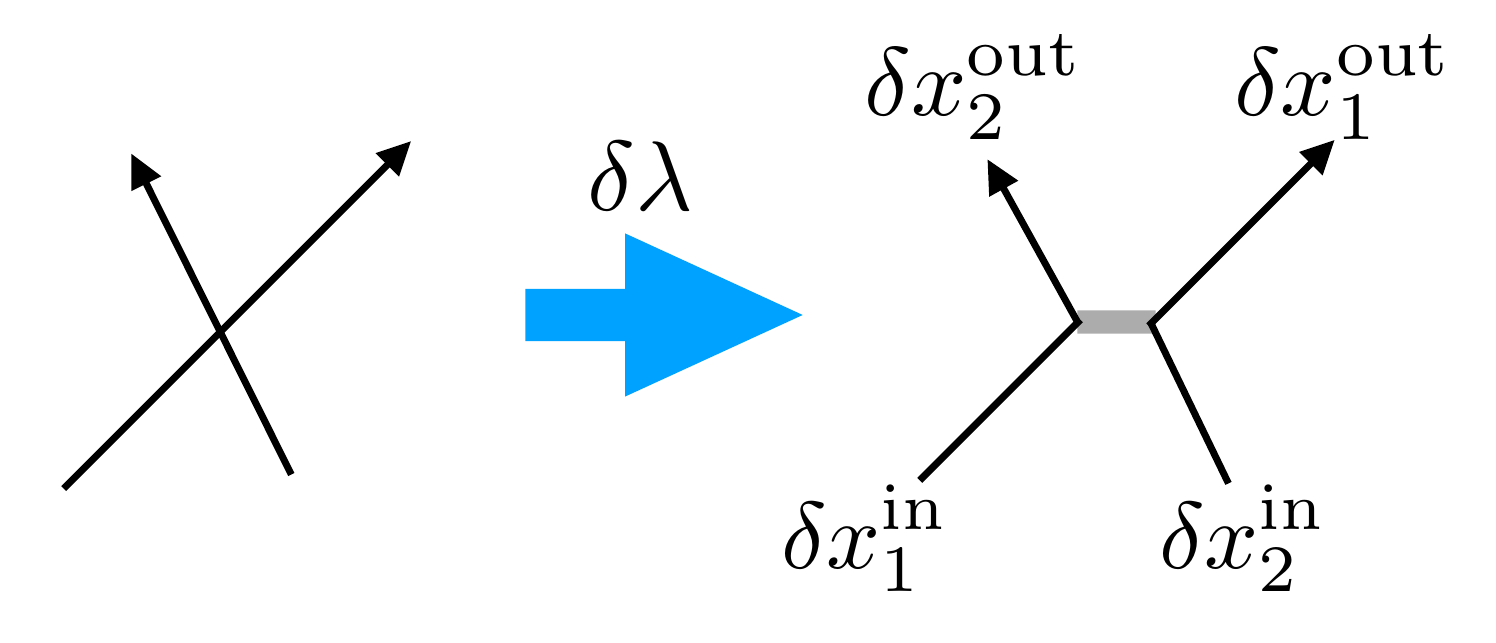}
\ec
\caption{The effect of the $\lambda$-deformation on a two-body scattering process. This illustration, where the fundamental particle gains a positive width $\delta\lambda$, corresponds to a negative scattering shift (or differential scattering phase)  $\varphi$, in the convention of \cite{DYC18}.}
\label{figscat}
\end{figure}

Let us apply an infinitesimal transformation $\delta\lambda$. We are interested in the trajectories under $H^{(\lambda)}$, which are have positions $x^{(-\lambda)}_n(t)$. Using \eqref{transfo}, before the collision, the $\lambda$-displacements are therefore $\delta x_1^{\rm in} = 0$, $\delta x_2^{\rm in} = \delta\lambda$, and after collision, once particles have crossed, $\delta x_2^{\rm out} = 0$, $\delta x_1^{\rm out} = \lambda$. The total changes are therefore (see Fig.~\ref{figscat})
\beq\label{phi12}
	\delta\varphi_{1} = - \delta\varphi_{2} =
	-\delta\lambda.
\eeq
In the quantum realm, this is associated with a scattering angle as per \eqref{scatphase},  and we obtain the Galilean invariant deformation
\beq\label{Phi12}
	\delta\Phi(p_1,p_2) = \delta \lambda \,(p_2-p_1).
\eeq

If the undeformed system is integrable, then so is the deformed one, and the above results are extended to multi-particle scattering by factorised scattering; the full deformation of the scattering angle is
\beq
	\delta\Phi(p_1,\ldots,p_N) = \sum_{m<n}
	\delta \lambda\,(p_n-p_m).
\eeq

If the undeformed system is not integrable, then the momenta are not preserved, and we cannot follow individual momenta. Two sets of momenta are necessary to describe the scattering: $p^{(\rm in)}_n$ and $p^{(\rm out)}_n$, and for clarity, we keep the labelling corresponding to individual particles for both sets of momenta. The full permutation
\beq\label{fullsigma}
	\sigma(n) = N+1-n
\eeq
then insures that the same numerical ordering holds in the in out regions: $p^{(\rm in)}_n>p^{(\rm in)}_{n+1}$, $p^{(\rm out)}_{\sigma(n)}>p^{(\rm out)}_{\sigma(n+1)}$.  We may work out the scattering deformation phase by realising that the deformation leads to constant shifts in both asymptotic regions, as per \eqref{psilambda}. These give rise to phase factors in the full S matrix, and opposite phases occur for in and out states (because the S matrix is a state overlap). Overall, the deformation of the S matrix may be written as in different ways:
\beq\label{Snontinteg}
	S \to e^{\ri\lambda \sum_{m<n} (p_n^{({\rm in})} - p_{\sigma(m)}^{({\rm out})})} S = e^{\ri\lambda \sum_{m<n} (p_n^{({\rm in})} - p_{n}^{({\rm out})})} S.
\eeq
The integrable results is recovered from the first expression using $p^{({\rm in})}_n = p^{({\rm out})}_{\sigma(n)}$.

This specialises naturally to the system of free hard rods, which is integrable and where the scattering shifts \eqref{phi12} and angle \eqref{Phi12} are discussed in \cite{DYC18}. As the scattering phase of free particles is simply the unit, this can be seen as the Galilean equivalent of the result of \cite{Zam2,Dub2,Tat1}. Indeed, in the latter work, it was found that the $T{\overline T}$ deformation, with, say, deformation parameter $\lambda_{\rm R}$, of an integrable model, modifies the scattering phase by the multiplication by a CDD factor $S(\theta-\alpha)= e^{\ri \lambda_{\rm R} p(\alpha-\theta)}$ where $\theta,\alpha$ are rapidities. 
In the Galilean case, since we find that (for $\lambda>0$) free particles are mapped onto hard rods, we therefore find a similar modification of the scattering: a multiplication by the phase
\beq\label{Shr}
	S(p-q) = e^{\ri \lambda (q-p)}.
\eeq

In the context of GHD, as mentioned, it was parallely understood that the hydrodynamics of integrable systems can be interpreted as that occurring under a change of metric induced by the free space associated to jump processes that are undergone by quasiparticles under interaction. In this interpretation, the jump lengths -- determining the change of the distance function -- are set to the differential scattering phase $\varphi=d \Phi/d p$ (and are in general momentum dependent) \cite{DYC18}. GHD therefore provides a quite general context for the geometric interpretation of the CDD factor seen from the $T{\overline T}$ deformation.

\subsection{Hydrodynamics}

Using the idea of a change of metric, it is a simple matter to extract the {\em hydrodynamic equations} for the $\lambda$-deformed system, once those for the undeformed system are known. This follows the techniques established in the context of GHD \cite{DSYgeom}.

Assume that the system $H$ admits a set of conserved charges with local or quasi-local densities, $Q_i = \int \dd x\,q_i(x)$. Then the Euler-scale hydrodynamic equations are obtained by relating the average densities $\mathsf q_i = \bra q_i\ket$ and currents $\mathsf j_i = \bra j_i\ket$ by writing the continuity equations \eqref{cons} under the assumption of local maximisation of entropy,
\beq\label{hydro}
	\p_t \mathsf q_i + \p_x \mathsf j_i = 0.
\eeq
The condition of maximisation of entropy is that according to which at every space-time point, the state $\bra \cdots\ket$ is a (generalised) Gibbs ensemble, with density matrix $\h\rho(x,t)\propto e^{-\sum_i \beta^i(x,t) Q_i}$. Eq.~\eqref{hydro} holds at large scales in space and time; see e.g.~\cite{Spohn-book,DoyonLecture2020}.

Recall that the particle density $\rho(x) = \mathsf q_0(x)$ is conserved. Following the usual convention, we may write the current as $\mathsf j_0(x) = v(x)\rho(x)$, so that
\beq\label{standard1}
	\p_t \rho + \p_x (v\rho) = 0.
\eeq

In Galilean invariant models (that is, with $\omega(p) = p^2/(2M)$), the particle current is itself a conserved density, the momentum density, $\mathsf j_0=\mathsf q_1$. Consider the simple case where the fluid is non-integrable -- so that only the particle number, momentum and energy are conserved. Further, consider the fluid to be isentropic -- so that the hydrodynamic equation for energy may be neglected; this is in many cases a very good approximation. Under such conditions, the two hydrodynamic equations for conservation of particles and momentum take the form \eqref{standard1} and
\beq\label{standard2}
	\p_t v + v\p_x v = -\frc{\p_x \mathsf P(\rho)}{\rho} 
\eeq
where $\mathsf P(\rho)$ is the equilibrium pressure, as a function of the density. The pressure function is the ``equation of state" which encodes the specific non-integrable, Galilean model (that is, the potential $V$ in \eqref{H}). In general -- in integrable models, without Galilean invariance, etc. -- one must instead consider the full set of equations \eqref{hydro}, for all conserved densities $\mathsf q_i$.

We are looking for the hydrodynamics equations for the deformed system $H^{(\lambda)}$. According to \eqref{transfo}, this can be obtained by ``expanding" the particles' widths (here and below we use the language corresponding to $\lambda>0$, but the same equations hold for $\lambda<0$); or geometrically, by using the metric \eqref{metric}. Consider coordinate $\t x$ along which the elongated particles lie, and their density $\t\rho(\t x,t)$. Then we have
\beq\label{trho}
	\t\rho(\t x,t)\, \dd \t x = \rho(x,t)\, \dd x
\eeq
and from \eqref{metric}, the change of metric
\beq\label{metricrho}
	\dd \t x = (1+\lambda \rho(x,t))\, \dd x\quad \Leftrightarrow\quad
	\dd x = (1-\lambda \t \rho(\t x,t)) \,\dd  \t x.
\eeq
As discussed in \cite{} in the context of the GHD of hard rods, the quantity $\rho_{\rm s}(\t x,t) = 1-\lambda \t \rho(\t x,t)$, measuring the free space density, can be identified with the ``density of state" from the thermodynamic Bethe ansatz. Likewise, $1+\lambda \rho(x,t)$ is the added-width density. Importantly, both satisfy a continuity equation. For the former, this will follow from the result we derive below;   for the latter, it follows from \eqref{standard1}:
\beq\label{consstate}
	\p_t \rho_{\rm s}(x,t) +\lambda \p_x (v\rho(x,t)) = 0.
\eeq

Below, in order to lighten the notation, tilde-quantities are evaluated at $(\t x,t)$ and non-tilde at $(x,t)$, under the relation \eqref{metricrho}. Let us define the following conserved densities and currents in the $\lambda$-deformed system:
\beq
	\t{\mathsf q}_i \,\dd\t x = \mathsf q_i \,\dd x,\quad
	\t {\mathsf j}_i = \mathsf j_i - \lambda v\rho \,\t {\mathsf q}_i.
\eeq
The first equation of course specialises to \eqref{trho} at $i=0$, representing the density in the $\lambda$-deformed system. The latter equation, with $\t{\mathsf j}_0 = \t v \t \rho$, gives $\t v = v$. Then, using the chain-rule relation
\beq
	\p_t\big|_{\t x} = \p_t\big|_x + \lambda v \rho\,\p_{\t x}\big|_t
\eeq
which follows from \eqref{metricrho} and \eqref{consstate}, one can show that \eqref{hydro} implies
\beq
	\p_t \t{\mathsf q}_i + \p_{\t x} \t{\mathsf j}_i = 0.
\eeq
These are the hydrodynamic equations for the $\lambda$-deformed system. They contain in particular the conservation equation
\beq\label{lstandard1}
	\p_t \t\rho + \p_{\t x} (\t v \t \rho) = 0
\eeq
for the density of $\lambda$-elongated particles.

In the isentropic, Galilean invariant case, where only \eqref{standard1} and \eqref{standard2} remain, in the deformed system we have \eqref{lstandard1} and
\beq
	\p_t \t v + v \p_{\t x} v = -\frc{\p_{\t x} \t{\mathsf P}(\t \rho)}{\t\rho}
\eeq
where the new pressure is simply obtained form $\mathsf P(\rho)$ by the transformation of densities,
\beq
	\t{\mathsf P}(\t \rho) = \mathsf P\Big(\frc{\t \rho}{1-\lambda \t \rho}\Big).
\eeq
In this case, the deformed system is still Galilean and isentropic, but the pressure is modified by the change of metric.

\subsection{Thermodynamics}

It is interesting to discuss the effect of the $\lambda$-deformation on the thermodynamics of the system. In order to do so, the system is put on a finite region (say with hard-wall boundary conditions) of length $R$, and, in the thermodynamic limit, $N,R\to\infty$ with fixed density $\rho = N/R$. We assume again the potential to be decaying  fast enough at large separations. For the present discussion, we also assume that it is well behaved at short distances (apart from the hard-core condition). The change-of-metric interpretation immediately gives us the main features of the thermodynamics of the system $H^{(\lambda)}$. For $\lambda>0$, the free space is reduced, and thus the volume on which the particles can effectively travel is $R^{(\lambda)} = R-\lambda N$. As a consequence, there is a maximum density $\rho_{\rm max}^{(\lambda)} = \lambda^{-1}$ in the undeformed system, for the deformation $\lambda$ to be applicable. As the maximal density is approached, because the potential is well behaved at short distances, the interaction can be neglected. Thus we have a system of effectively free particles on an effectively small region. Therefore, in the quantum case, as the maximal density is approached, all energies tend to infinity like $(R-\lambda N)^{-2}$: each deformed energy is related to its undeformed value by (see \eqref{EE0nonrel})
\beq\label{ER}
(R-\lambda N)^2E^{(\lambda)} =R^2E^{(0)}.
\eeq
Thus for a fixed particle density $N/R$, the density of eigenstates is very similar to that of the undeformed free fermion (or hard core boson) system, growing in a conventional manner at large energy density. However, if we fix instead the energy per particle $E^{(\lambda)}/N$, the left-hand side of (\ref{ER}) has a maximum as a function of $E^{(\lambda)}$, as does the density of states. This is then very similar to the relativistic case, where, in the canonical ensemble the maximum energy corresponds to infinite temperature, and energies above this to negative temperatures. Such phenomena occur, for example, in spin systems, and in general where the phase space per degree of freedom is finite. In this case it occurs because of the maximum density implying a maximum energy. However it should be noted that the maximum thermodynamic energy is less than the maximum of the left hand side of (\ref{ER}), because of entropic effects. 

On the other hand, for $\lambda<0$, the effective volume grows as like $R+|\lambda| N$, and the mean spacing between the particles is proportional to $(R/N)+|\lambda|$. Thus for $|\lambda|$ large, because of the finite range of the potential the particles can again be considered free. Moreover, as this is an effectively low-density situation, we may use Boltzmann statistics. In that case, in the grand canonical ensemble, with fugacity $z$, the partition function is
\beq\label{GCE}
\Omega=\sum_{N=0}^\infty (z^N/N!)Z_1(R+|\lambda| N)^N
\eeq
where $Z_1(R)$ is the single-particle partition function in volume $R$. At inverse temperature $\beta$, we have $Z_1(R)=(R/2\pi)\int e^{-\beta p^2/2M}dp= R\sqrt{M/(2\pi \beta)}$, and by the usual application of Stirling's formula, as the sum is dominated by terms with $N$ large, we have
\beq
	\sum_{N=0}^{\infty} \big(ez|\lambda|\sqrt{M/(2\pi\beta)} \big)^N /\sqrt{N}
\eeq
which diverges for temperatures
\beq\label{betastar}
	\beta^{-1}> \beta^{-1}_* ,\quad \beta_* = \frc{ Me^2 z^2|\lambda|^2}{2\pi}.
\eeq
As a consequence, the free energy has a square root singularity there. In fact, an exact calculation of the free energy in the free hard-rod model has been done in \cite[App D]{Myers-Bhaseen-Harris-Doyon}, using the thermodynamic Bethe ansatz, giving (the result is valid for any sign of $\lambda$)
\beq
	f = - \sqrt{\frc{M}{2\pi\beta}} \exp\Big[-W\Big(\lambda\sqrt{M/(2\pi \beta)}\Big)\Big]
\eeq
where $W$ is the Lambert $W$ function on its principal branch. As $W(z) + 1 \sim \sqrt{2e^{-1}(z+e^{-1})}$ as $z\to -e^{-1}$, the free energy, although finite, indeed has a square-root behaviour and becomes complex for $\lambda<0$ at $\beta = \beta_*$. There is therefore a maximum temperature, due, in this case, to the expansion of the phase space. This is the non-relativistic analogue of the Hagedorn transition, already  noted for the $T\b T$ deformation of relativistic theories \cite{Tat1} and already known to be a feature of the world sheet description of the Nambu-Goto string \cite{Cas}\footnote{We note that the Hagedorn-type behaviour happens here for $\lambda<0$, while in most of the literature on relativistic $T\overline T$-deformations it is found for $\lambda>0$. The reason for this is that the relevant quantity is the product $E_0\lambda$, where $E_0$ is the ground state energy. In the non-relativistic case we take $E_0>0$ to be the (positive) rest mass energy -- equivalently, we take a negative chemical potential -- while for relativistic models one normally subtracts off the divergent bulk ground state energy, leaving the (negative) finite-size Casimir energy $E_0(R)<0$.}.

\subsection{Many species}\label{ssectmanyspecies}

It is finally instructive to consider the case where the system is composed of many species of non-identical particles. Each species could, for example, have a different kinetic mass, so that the momentum-velocity relationship would depend on this.

Again, we will consider two separate physical situations: the hard-core picture, and the go-through picture, see Section \ref{ssectinteracting}, the discussion after Eq.~\eqref{translationinv}. Here, because of the presence of many species, the two pictures are not simply related by a mapping of trajectories, and lead to quite different deformations.

Let us start with the hard-core picture, where particles stay ordered on the line, with coordinates $x_n$. We may then think of each carrying its own effective deformation parameter $\lambda_n$. To be more precise, we construct the conserved densities
\beq\label{q0n}
	q_{0,n}(x,t) = \delta(x_n(t)-x),
\eeq
and the deformation has the form $H^{(\{\lambda\})}\to H^{(\{\lambda+\delta\lambda\})} = H^{(\{\lambda\})}+\delta\lambda\Delta^{(\{\lambda\})} + O((\delta\lambda)^2)$ with
\beq\label{Deltan}
	\Delta^{(\{\lambda\})} = \sum_n\ell_n \int \dd x\,(q^{(\{\lambda\})}_{0,n}(x-\varep)j^{(\{\lambda\})}_1(x)- j^{(\{\lambda\})}_{0,n}(x-\varep)q^{(\{\lambda\})}_1(x) )
\eeq
where we set $\lambda_n = \lambda \ell_n$ and $\delta \lambda_n = \delta \lambda \ell_n$. This can be obtained from the deformation generator
\beq\label{Xgenn}
	X = -\sum_n \ell_n \int \dd y\,\int_{-\infty}^{y-\varep} \dd x\,q_{0,n}(x)q_1(y)
\eeq
with the flow equation
\beq
	\frc{d H^{(\{\lambda\})}}{d\lambda} = \{H^{(\{\lambda\})},X\}.
\eeq

Contrary to the case where all deformation parameters are the same, here the dependence of some physical quantities on the deformation parameters is affected by the choice of a point of anchor: the exact meaning of the coordinate $x_n$ for the $\lambda$-elongated particle $n$ (say with $\lambda_n>0$). In fact, this point of anchor is related to the choice of the point-splitting infinitesimal $\varep$ in \eqref{Deltan}. For $\varep>0$ as chosen, it is the left-most point of the rod; for $\varep<0$ it would be the right-most point; while for $\varep=0$, with the usual regularisation of the step function implicit in \eqref{Xgenn}, it would be the middle point. Let us choose, as above, $\varep>0$: the left-most point.

The generalisations of \eqref{transfo} and (\ref{psilambda}) are then
\beq\label{transfon}
	p_n^{(\lambda_1,\lambda_2,\ldots)} = p_n,\qquad x_n^{(\lambda_1,\lambda_2,\ldots)} = x_n - \sum_{j=1}^{n-1}\lambda_j.
\eeq
and
\beq
\psi^{(\lambda_1,\lambda_2,\ldots)}(x_1,\ldots,x_N)=\psi\big(x_1,x_2-\lambda_1,x_3-\lambda_1-\lambda_2,\ldots,x_N-\sum_{n=1}^{N-1}\lambda_j\big).
\eeq
Recall that in the hard-core picture, the non-transmission condition remains valid in the quantum case (see the discussion in the paragraph after Eq.~\eqref{EE0nonrel}).

Accounting for displacements of particles in the in and out asymptotic regions, the deformation CDD factor in the scattering amplitude is then
\begin{eqnarray}
S&\to &e^{\ri\sum_{n=1}^N(\lambda_1+\cdots+\lambda_{n-1})(p_n^{(\rm in)}-p_{n}^{(\rm out)})}S\n
&=&e^{\ri\sum_{m<n}\lambda_m(p_n^{({\rm in})}-p_{n}^{({\rm out})})}S 
\qquad\qquad\mbox{(hard-core particles)}\label{manyspecies}
\end{eqnarray}
which generalises one form of \eqref{Snontinteg}. Using the permutation \eqref{fullsigma}, under which the natural ordering $p^{(\rm out)}_{\sigma(n)}>p^{(\rm out)}_{\sigma(n+1)}$ holds, this becomes
\begin{eqnarray}
S&\to &e^{\ri\sum_{n=1}^N((\lambda_1+\cdots+\lambda_{n-1})p_n^{(\rm in)}-(\lambda_{\sigma(n+1)}+\ldots+\lambda_{\sigma(N)})p_{\sigma(n)}^{(\rm out)})}S\n
&=&e^{\ri\sum_{m<n}(\lambda_mp_n^{({\rm in})}-\lambda_{\sigma(n)}p_{\sigma(m)}^{({\rm out})})}S,\label{manyspecies2}
\end{eqnarray}
generalising the other form of \eqref{Snontinteg}. We note that this agrees with the non-relativistic limit of \cite[Eq 1.1]{Dub2}; there it is argued that the CDD dressing factor is valid for any undeformed S-matrix in two dimensions, integrable or not, and where particles may for instance have different masses.

It is interesting to note what happens if the original model, whence also the deformed model, is integrable. Let us restrict to the cases where the original particles are the correct asymptotic particles (that is, there are no bound states). In this case momenta are preserved\footnote{This excludes the case of particles with different masses, as 2-body scattering of hard-core particles with different masses does not preserve their momenta.}, and therefore $p_{\sigma(n)}^{({\rm out})} = p_{n}^{({\rm in})} \equiv p_{n}$ as momenta must be fully permuted (Eq.~\eqref{fullsigma}). It is important to remark that the undeformed scattering $S$ is {\em not diagonal}, as the hard-core condition imposes that particle types be exchanged. In terms of a Zamolodchikov-Faddeev algebra, it takes the simple form
\beq\label{reflectionundeformed}
	Z_j(p)Z_{j+1}(q) = Z_j(q) Z_{j+1}(p)
\eeq
where, because the ordering is always kept, only consecutive particle types, $j$, $j+1$, can ever interact with each other.

Likewise, the resulting deformation factor
\beq
	e^{\ri\sum_{m<n}\lambda_m(p_n-p_{\sigma(n)})}
	\qquad\qquad\mbox{(hard-core particles, integrable)}\label{manyspeciesintegrable}
\eeq
from \eqref{manyspecies} is {\em not} a product of two-body scattering deformation factors.  Again this is because with hard-core particles, species get reflected at collisions, instead of transmitted, as the ordering of species stays unchanged in space while momenta are transmitted.  For instance, the two-body case $e^{\ri( \lambda_1 p_2 - \lambda_1 p_1)}$ reflects the choice of the left-most point of the rod as its coordinate, which implies that the scattering of velocity tracer $n$ with $n-1$ on its left induces a jump by $\lambda_{n-1}$, while with $n+1$ on its right induces a jump by $\lambda_n$. However, although such jumps stay true in two-body sub-events of many-body scattering, the connection momentum-species is broken.

Eq.~\eqref{manyspeciesintegrable} gives then a simple example of non-diagonal scattering generalising \eqref{reflectionundeformed}. The deformation factor is a purely-reflective solution to the Yang-Baxter equations. This can be seen explicitly by writing the Zamolodchikov-Faddeev algebra:
\beq
	Z_j(p) Z_{j+1}(q) = e^{\ri \lambda_j (q-p)} Z_j(q)Z_{j+1}(p).
\eeq
One can verify that this is associative, by evaluating
\beq
	Z_1(p) Z_2(q) Z_3(r) = e^{\ri (\lambda_1+\lambda_2)(r-p)}Z_1(r)Z_2(q)Z_3(p)
\eeq
following the two different exchange paths:
\beq
\lambda_1 (q-p) + \lambda_2(r-p) + \lambda_1(r-q)
=
\lambda_2(r-q) + \lambda_1(r-p) + \lambda_2(q-p)
=
(\lambda_1+\lambda_2) (r-p)
\eeq

\begin{figure}
\bc
\includegraphics[width=4cm]{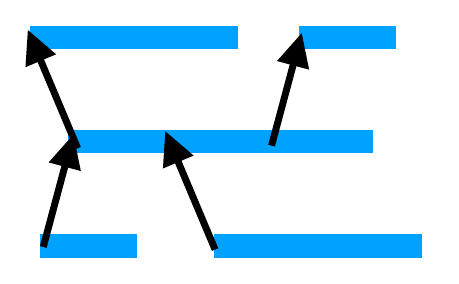}
\ec
\caption{How enlarged particles collide in the go-through picture after a many-species $\lambda$-deformation.}
\label{figgothrough}
\end{figure}

Let us now consider the go-through picture. With many species, the simple relation between the hard-core and go-through pictures obtained by permutation of particles does not hold anymore. Indeed, in the go-through picture, particles species would be transmitted instead of reflected; thus a simple permutation cannot account for the difference in dynamics. The direct construction of the deformation without hard-cord condition, where the connection momentum-species stays unbroken at collisions, is more delicate. Indeed, the breaking of ordering breaks the argument leading to the simple shift, whenever particles are not well separated enough. Without going into the details of the resulting calculations, it is natural to expect that at collisions, particles of species $n$ and $m$ (which are not necessarily ordered in space anymore) instantaneously ``exchange their lengths" in addition to exchanging their momenta, by coming together momentarily into a new particle of length $\lambda_n+\lambda_m$, and immediately re-dividing appropriately. It is simple to see that this gives rise to a consistent microscopic dynamics. See Fig.~\ref{figgothrough}.

Despite the absence of a clear microscopic derivation of the collision physics in the go-through picture in the many-species case, it is in fact possible to deduce the resulting scattering phase deformation without hard-core condition. The scattering phase deformation only requires understanding the effects of the $X$ generators in asymptotic regions, where particles are well separated and the calculation we performed above works without modification. Take, then, a model of $N$ particles of different species, with coordinates $x_n$ and momenta $p_n$. Coordinates $x_n$ are not always ordered, as this is the go-through picture. They are assumed to be ordered in the incoming region, $x_n<x_{n+1}$ as $t\to-\infty$, with momenta $p_n^{({\rm in})}>p_{n+1}^{({\rm in})}$. Scattering leads to a generically nontrivial permutation $\sigma'$ such that $p_{\sigma'(n)}^{({\rm out})}>p_{\sigma'(n+1)}^{({\rm out})}$.  The precise element $\sigma'$ depends on the momenta $p^{({\rm in})}_n$ and on the interaction in an intricate way. But whatever it is, the resulting deformation is
\begin{eqnarray}
S&\to &e^{\ri\sum_{n=1}^N( (\lambda_1+\cdots+\lambda_{n-1})p_n^{(\rm in)}-(\lambda_{\sigma'(n+1)}+\cdots+\lambda_{\sigma'(N)})p_{\sigma'(n)}^{(\rm out)})}S\n
&=&e^{\ri\sum_{m<n}(\lambda_mp_n^{({\rm in})}-\lambda_{\sigma'(n)} p_{\sigma'(m)}^{({\rm out})})}S
	\qquad\qquad\mbox{(go-through particles)}.
\label{manyspeciesthroughgen}
\end{eqnarray}
Note that despite the similarity with \eqref{manyspecies2}, this is generically a different scattering shift, as $\sigma'$ depends on momenta and interaction in a nontrivial fashion.

Again let us consider the case of an original integrable model, and further assume that the permutation element is trivial, $\sigma' = {\rm id}$. Because this is a go-through picture, this means that particles fully exchange their order in any scattering event -- for instance, this holds if the original model is a free particle system.  Then, with $p_n = p_n^{({\rm in})} = p_n^{({\rm out})}$ the deformation reduces to
\begin{eqnarray}
S&\to &e^{\ri\sum_{m<n}(\lambda_mp_n-\lambda_n p_{m})}S
	\qquad\qquad\mbox{(go-through particles, integrable).}
\label{manyspeciesthroughintegrable}
\end{eqnarray}
Note the difference with \eqref{manyspeciesintegrable}. Here the deformation factor is clearly a product of two-body scattering deformation factors, in agreement with the fact that scattering is purely transmissive and thus diagonal. Specifically, the term, in the sum, with positive (negative) sign corresponds to the scattering of velocity tracer $n$ ($m$) with a particle on its left (right).


\section{Relativistic case} \label{sectrel}

The interpretation of the $qj$ deformation as an elongation of the fundamental particles of the theory carries over to the relativistic case. Recall that in the non-relativistic limit, the energy-momentum relation is approximated to $E=\sqrt{M^2c^4+c^2p^2}\sim Mc^2+p^2/2m+\cdots$. Correspondingly, the relativistic energy 2-current breaks up into two separately conserved pieces
\beq
T^{0a}\to Mc^2j^a+T^{0a}_{\rm NR}
\eeq
where $j^a = (q_0,j_0)^a$ is the particle 2-current as before, and $T^{0a}_{\rm NR}$ is the non-relativistic energy 2-current. Therefore one way of viewing the previous results is to treat only the rest mass term as being dominant, in which case the deformation parameter $\lambda$ we have been using is related to the relativistic parameter $\lambda_{\rm R}$ of the $\lambda_{\rm R} T\b T$ deformation by
\beq
\lambda=\lambda_{\rm R}Mc^2
\eeq
Obviously this identification works only for massive states (or massive particles) in the relativistic theory. However, this suggests an immediate relativistic generalisation of particle widening: under the relativistic $T\b T$ deformation,
\begin{center}
\em Each particle of rest mass $M$ acquires a spatial width $\lambda_{\rm R}Mc^2$ in its own rest frame.
\em
\end{center}
In a boosted frame, where it has energy-momentum $(p^0,p^1)$, it therefore acquires a spatial width $\lambda_{\rm R}p^0$, while the end-point of the hard rod is boosted to a different time $\lambda_{\rm R}p^1$ relative to the centre its start-point (the combination of these effects lead to the usual relativistic length contraction). In what follows, we set
\beq
	c=1.
\eeq

\begin{figure}
\bc
a. \includegraphics[width=7cm]{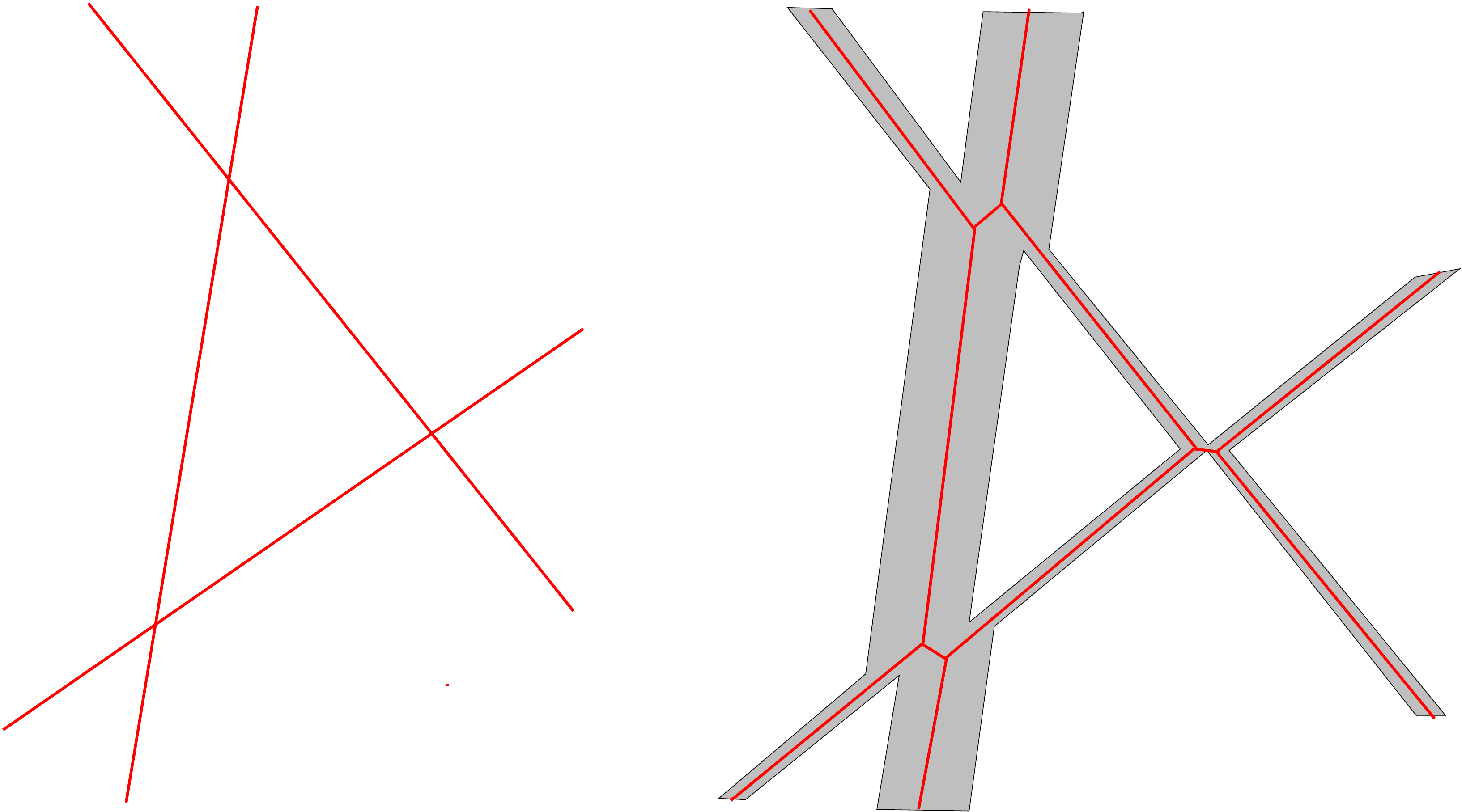}\qquad b.
\includegraphics[width=7cm]{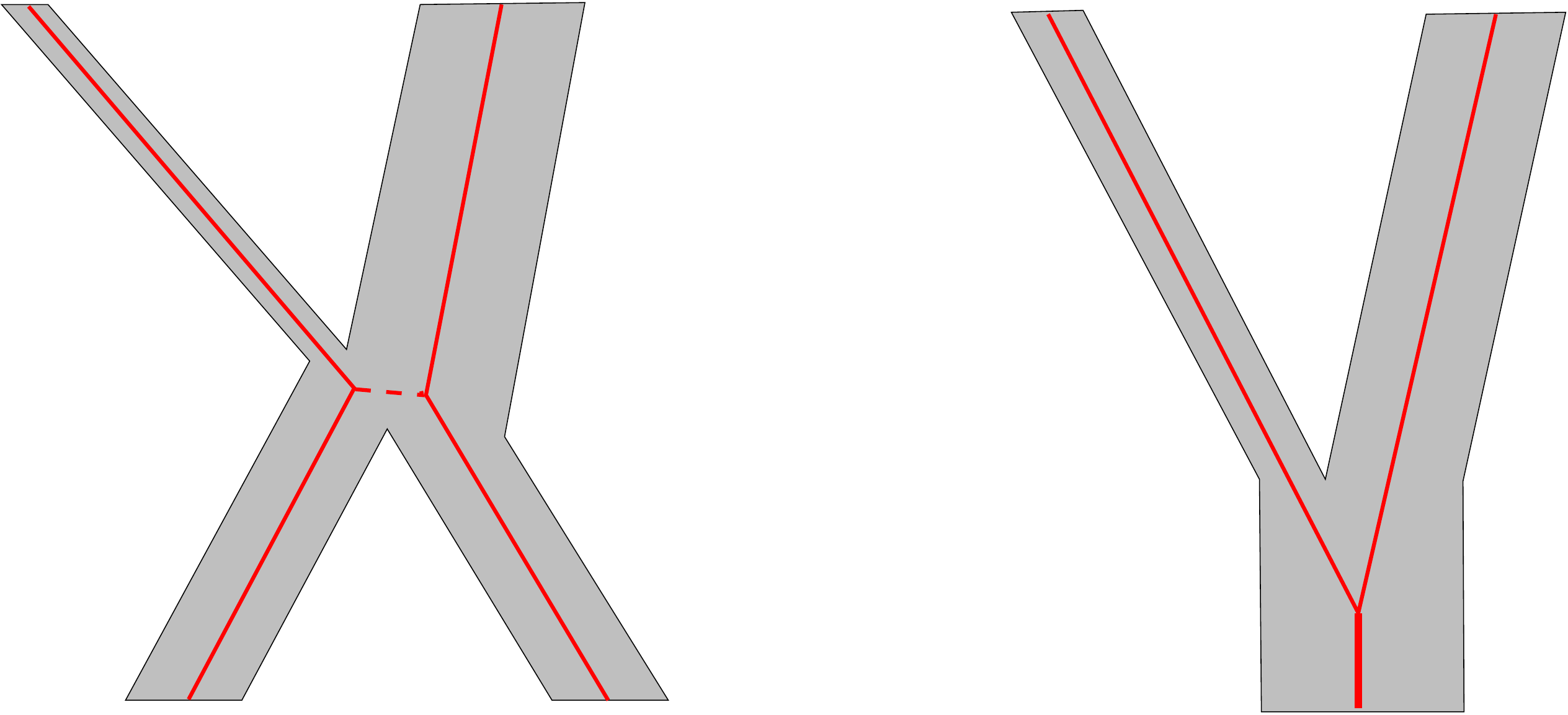}
\ec
\caption{The effect of the $\lambda_{\rm R}$-deformation on relativistic scattering processes. The particles gain a width in a  consistent fashion in space-time, as if ``grout" were added between tiles. a. elastic scattering; b. inelastic scattering.}
\label{figgrout}
\end{figure}

This picture is consistent with both elastic an inelastic scattering processes: since in any such process $\sum_ip^0_i$ and $\sum_ip^1_i$ are conserved, so is the total width and the relative time delay between its ends. For successive scattering processes this leads to a consistent polygonal tiling of Minkowski space, with the insertion of `grout' between the tiles. This is illustrated in Fig.~\ref{figgrout}. 

Moreover, it implies that \em any \em state of definite total energy $E$ and zero momentum acquires an additional width $\lambda_{\rm R}E$, so that if the system had size $R$ before deformation, the effective size, after deformation, is $R-\lambda_{\rm R} E$. Therefore the deformed energy eigenvalues in finite size $R$ obey (this is the relativistic generalisation of \eqref{EE0nonrel})
\beq
E^{(\lambda_{\rm R})}(R)=E^0\big(R+\lambda_{\rm R} E^{(\lambda_{\rm R})}(R)\big)
\eeq
which \cite{Zam2} is the implicit solution to the inviscid Burgers' equation
\beq\label{Be}
\partial_{\lambda_{\rm R}} E^{(\lambda_{\rm R})}(R)=E^{(\lambda_{\rm R})}(R)\partial_RE^{(\lambda_{\rm R})}(R)
\eeq
as derived by Zamolodchikov \cite{Zam1,Zam2}. In fact the simple linear equation $\partial_{\lambda_{\rm R}} R=-E$ arises on solving (\ref{Be}) by the method of characteristics. 

Lastly, this picture gives a simple explanation of the relativistic CDD factors. We may take inspiration from the multi-specie non-relativistic formula \eqref{manyspeciesthroughintegrable}, where the term, in the sum with, with positive (negative) sign corresponds to the scattering of velocity tracer $n$ ($m$) with a particle on its left (right). In the relativistic case, the scattering of particle $n$ with $m\neq n$ on either side induces a jump by $\lambda_{\rm R} p^0_m$, and we obtain
\beq\label{phaserela}
e^{\ri \lambda_{\rm R}\sum_{m<n}(p^0_mp^1_n-p^0_np^1_m)}
\eeq
as found in \cite{Dub2,Zam2}.

It is a simple matter to derive this picture in the case of free relativistic particles by performing an analysis paralleling that of non-relativistic systems in Section \ref{sectmain}. We simply choose the relativistic dispersion relation
\beq
	\omega(p) = \sqrt{M^2 + p^2}
\eeq
in \eqref{H}, and, in order to have Lorentz invariance, we set the potential to zero whenever particles are away from each other. The deformation is generated by the $X$ operator \eqref{Xgen}, but with the energy density $i=2$ instead of the particle density $i=0$, where
\beq\label{q0q1rel}
	q_2(x,t) = \sum_n \delta(x_n(t)-x)E_n(t)
\eeq
with $E_n = \omega(p_n)$. Again, the full deformation of the Hamiltonian is the result of the flow $\lambda_{\rm R} \mapsto H^{(\lambda_{\rm R})}$ generated by $X$,
\beq\label{flow}
	\frc{d H^{(\lambda_{\rm R})}}{d\lambda_{\rm R}} = \{H^{(\lambda_{\rm R})},X\}.
\eeq
The generator $X$ may again be explicitly worked out:
\beqa
	X &=& -\sum_{n,m} \int \dd y\int_{-\infty}^{y-\varep}\dd x\,\delta(x-x_n) \delta(y-x_m) E_n p_m\n
	&=& -\sum_{n,m} \int_{-\infty}^{x_m-\varep}\dd x\,\delta(x-x_n)E_np_m  = -\sum_{n,m} \Theta(x_m-\varep-x_n)E_np_m.
\eeqa
Clearly, when particles are non-coincident, the momenta do not flow, $\{p_n,X\} = 0$, and the particle coordinates flow as
\beq\label{reldis}
	\{x_n,X\} = -\sum_{x_m<x_n} E_m - v_n \sum_{x_m>x_n} p_m
\eeq
where $v_n = \omega'(p_n) = p_n/E_n$ is the relativistic velocity of particle $n$.

One way of extracting the physical meaning of this flow is to go to the frame in which the total momentum of the rest of the particles $m\neq n$ is boosted to zero. In particular, if there are only two particles in total, the single particle $m\neq n$ is therefore in its rest frame, where its width is to increase by $\lambda$. It this frame, we find
\beqa
	\{x_n,X\}
	 &=& -\sum_{x_m<x_n} (E_m-v_np_m) \n
	 &=& -\sum_{x_m<x_n} \frc{\cosh(\theta_m-\theta_n)}{\cosh(\theta_n)}
\eeqa
where $\theta_m$'s are the particles' rapidities. As the trajectories of the deformed systems have positions $x^{(-\lambda)}_n$ (see  \eqref{HHl} the discussion after \eqref{Hlambda}), for a small $\delta \lambda$ the deformed trajectories have displacements
\beq
	\delta x_n = \delta \lambda \sum_{x_m<x_n} \frc{\cosh(\theta_m-\theta_n)}{\cosh(\theta_n)}.\label{disspecial}
\eeq

The interpretation of this formula is as follows. Consider a small deformation by $\delta \lambda>0$, corresponding to an increase of the width of the particles; and a particle $m$ with $x_m<x_n$. In the rest frame of particle $m$, it elongates by $\delta\lambda = \delta\lambda_{\rm R} M$, and thus it observes a displacement of particle $n$ by a distance $\delta\lambda$. By contraction of lengths, if $\Delta_{n|m}^{\rm rest}$ is the corresponding displacement of particle $n$ in its own rest-frame, this means $\delta\lambda = \Delta_{n|m}^{\rm rest} /\cosh(\theta_m-\theta_n)$. In the laboratory frame, the observed displacement of particle $n$, again by contraction of lengths, is $\Delta_{n|m}^{\rm lab} = \Delta_{n|m}^{\rm rest}/\cosh(\theta_n)$. Thus indeed
\beq
	\Delta_{n|m}^{\rm lab} = \delta\lambda \frc{\cosh(\theta_m-\theta_n)}{\cosh(\theta_n)}
\eeq
in agreement with \eqref{disspecial}.

The resulting deformation of the two-body scattering shift may also be evaluated. Consider the scattering of velocity tracers 1 and 2, with rapidities $\theta_1$ and $\theta_2$, with $x_1<x_2$ in the ``in" configuration and $x_1>x_2$ in the ``out" configuration. Using \eqref{reldis}, before the collision, the displacements are $\delta x_1^{\rm in} = v_1p_2\delta\lambda$, $\delta x_2^{\rm in} = E_1\delta\lambda$, and after collision, once particles have crossed, $\delta x_2^{\rm out} = v_2p_1\delta\lambda$, $\delta x_1^{\rm out} = E_2\delta\lambda$. The scattering shift of velocity tracer 1 is therefore, from Eq.~\eqref{varphi1},
\beq
	\varphi_{1} = \delta x_1^{\rm in} - \delta x_1^{\rm out} = 
	-\delta \lambda\frc{\cosh(\theta_2-\theta_1)}{\cosh\theta_1}.
\eeq
This displacement is associated with a scattering phase $\Phi(\theta_1,\theta_2)=-\ri \,\log S(\theta_1,\theta_2)$ as per \eqref{scatphase}, giving the relativistically invariant phase
\beq
	\Phi(\theta_1,\theta_2) = \delta \lambda M\sinh(\theta_2-\theta_1)
	= \delta \lambda_{\rm R}(p_1^0p_2^1 - p_1^1 p_2^0).
\eeq
This agrees with \eqref{phaserela}  when specialised to the two-particle scattering process.

\section{Conclusion} \label{sectcon}

In this paper, we provided a simple geometric interpretation for the $T\overline T$ deformations of many-body 1+1-dimensional systems, both in the non-relativistic and relativistic cases. The deformations are seen as changing the widths of fundamental particles, elongating them for positive widths, and adding ``free space" for negative widths. The effects of the lengths of fundamental particles, including how this relates to scattering shifts and to a metric on space that depends on the state, have been studied in the context of GHD, the hydrodynamic theory for integrable systems. Independently, $T\overline T$ deformations have been seen to arise from field-dependent change of coordinates in relativistic quantum field theory. Our results provide a simple interpretation of the latter, and a new connection between $T\bar T$ deformations and ideas developed in GHD.

Our findings give an additional interpretation to the result according to which the Nambu-Goto action is generated by the flow \cite{Tat1}: world-sheet scattering arises from a stretching of the string, which corresponds to the addition of free space between the particles (negative widths) \cite{Dubovsky12}. We recall that other models implementing additional free space have been constructed \cite{DYC18,Donahue2019}, and that the concept of free space is closely related to that of the density of states in the TBA formulation of integrable thermodynamics (see e.g.~\cite{DoyonLecture2020}).

The result also gives an intuitive underpinning for the well studied fact that the $T{\overline T}$ deformation is irrelevant, and many of the features of the resulting theory are UV finite but non-local. Indeed, with elongated particles, there is a fundamental length, and it is not possible to describe configurations of particles at arbitrary small distances from each other.  The non-locality can also be seen in the differential scattering phase $\varphi$, which, at nonzero $\lambda$, does not vanish at large momentum differences -- a signal of the presence of a fundamental UV length.

In Subsection \ref{ssecthr}, we explained how the Galilean version of the $T\overline{T}$ deformation for positive deformation parameter $\lambda>0$ changes free particles to free hard rods, with rod lengths $\lambda$. For $\lambda<0$, the naive hard-rod interpretation does not work. However, it was noted in \cite{DYC18}\footnote{This originated from a private communication to BD from Herbert Spohn.} that the formal hard-rod model with negative rod lengths is related to the Lieb-Liniger model. Specifically, it gives the leading contribution in $1/c$ to the scattering, thermodynamics and hydrodynamics of the model, as the Lieb-Liniger coupling strength $c$ is made larger, with the identification $\lambda = -1/c$. Therefore, the Galilean $T\overline{T}$ deformation for negative $\lambda$ gives the Lieb-Liniger model at large coupling. Further discussions about the deformation of one-dimensional Bose gases can be found in \cite{jiang2020}. It would be interesting to investigate how further $1/c$ corrections can be obtained.

In recent works \cite{medenjak2020thermal,medenjak2020thermal2}, Medenjak, Policastro and Yoshimura studied nonequilibrium transport in the $T\overline T$ deformation of CFT's using GHD and the AdS/CFT correspondence, and universal results for the Drude weights were found. Interestingly, a relation was found between this deformation and the rule-54 cellular automaton, which is known to have a GHD description based on two particle types of equal and opposite bare velocities and of fixed negative widths. This therefore further confirms our findings.

As an additional connection between the literature on $T\overline T$ deformations and on GHD, we should mention the factorisation of expectation values of $T\overline T$ first obtained in \cite{Zam1}. For more general charge-current bilinears \cite{Zam2}, such a factorisation was obtained in the context of GHD in \cite[App B]{dNBD2}, from space-time translation invariance and clustering of the state. A direct consequence of this factorisation is the symmetry of the current susceptibility matrix, shown and discussed in various contexts \cite{toth2003onsager,grisi2011current,GHD1a,dNBD2,KarevskiCharge2019}. The symmetry of the current susceptibility matrix is a crucial ingredient for the existence of the entropy flux and the free energy flux in statistical mechanics (see the discussion in \cite{DoyonLecture2020}), and for the emerging hyperbolic system of hydrodynamic equations. It also gives rise to a proof \cite{yoshimura2020collision}, that applies to many integrable models, for the expression for exact average currents in GGEs first derived in \cite{GHD1a}.

Finally, it would be interesting to understand geometrically deformation by other currents. For instance, momentum-dependent differential scattering phases in integrable models are interpreted as momentum-dependent ``particle widths" \cite{DYC18}. One may naturally construct densities and currents associated to any given, fixed asymptotic momentum (which is conserved in integrable models), and thereby produce deformations giving rise to arbitrary integrable models. Our interpretation may also be useful to generalisations to higher dimensions, and may give a new viewpoint on gravity theories.

\medskip

\noindent {\bf\em Acknowledgment.} This work was made possible by the June 2018 International Conference on Mathematical Physics, Montr\'eal, Canada, where both authors were present, and our main idea -- that $T{\overline T}$-like deformations correspond to  extending the widths of fundamental particles -- emerged. BD also thanks B. Pozsgay for discussions, and T. Yoshimura for sharing the results of \cite{medenjak2020thermal} before publication. JC acknowledges support from the Quantum Science Center (QSC), at the University of California, Berkeley, a National Quantum Information Science Research Center of the U.S. Department of Energy (DOE), and from the Chern-Simons initiative also at the University of California, Berkeley. BD was supported in part by the Royal Society under a Leverhulme Trust Senior Research Fellowship, ``Emergent hydrodynamics in integrable systems: non-equilibrium theory", ref.~SRF\textbackslash R1\textbackslash 180103, and by the EPSRC under the grant ``Entanglement Measures, Twist Fields, and Partition Functions in Quantum Field Theory" ref.~EP/P006132/1.


\end{document}